# Predicted electronic markers for polytypes of LaOBiS$_2$ examined via angular resolved photoemission spectroscopy

Xiaoqing Zhou,[1] Qihang Liu,[1,*] J. A. Waugh,[1] Haoxiang Li,[1] T. Nummy,[1] Xiuwen Zhang,[1] Xiangde Zhu,[2] Gang Cao[3], Alex Zunger[1,*] and D. S. Dessau[1,*]

[1]University of Colorado at Boulder, CO 80504, USA

[2]High Magnetic Field Laboratory of the Chinese Academy of Sciences, Hefei, China

[3]University of Kentucky, Lexington, KY 40506, USA

[*]E-mail: qihang.liu85@gmail.com; daniel.dessau@colorado.edu; alex.zunger@colorado.edu.

**Abstract**

The natural periodic stacking of symmetry-inequivalent planes in layered compounds can lead to the formation of natural superlattices; albeit close in *total* energy, (thus in their thermodynamic stability), such polytype superlattices can exhibit different structural symmetries, thus have markedly different electronic properties which can in turn be used as "structural markers". We illustrate this general principle on the layered LaOBiS$_2$ compound where density-functional theory (DFT) calculations on the (BiS$_2$)/(LaO)/(BiS$_2$) polytype superlattices reveal both qualitatively and quantitatively distinct electronic structure markers associated with the Rashba physics, yet the *total* energies are only ~ 0.1 meV apart. This opens the exciting possibility of identifying subtle structural features via electronic markers. We show that the pattern of removal of band degeneracies in different polytypes by the different forms of symmetry breaking leads to new Rashba "mini gaps" with characteristic Rashba parameters that can be determined from spectroscopy, thereby narrowing down the physically possible polytypes. By identifying these distinct DFT-predicted fingerprints via ARPES measurements on LaBiOS$_2$ we found the dominant polytype with small amounts of mixtures of other polytypes. This conclusion, consistent with neutron scattering results, establishes ARPES detection of theoretically established electronic markers as a powerful tool to delineate energetically quasidegenerate polytypes.

**Introduction**

Ordered crystalline compounds of fixed composition are generally assumed to have a unique and specific crystallographic structure being distinctly separate from other phases in the low-temperature phase diagram. An exception is *polytypes* that represent an energetically closely spaced set of ordered compounds with the same composition often differing by orientations of certain sub-units. The best know example is zinc-blend and wurtzite polytypes of binary semiconductors[1] best exemplified by SiC that show ~100 polytypes[2], or ZnO, ZnS appearing each as either zinc-blend or wurtzite. What makes such polytypes electronically interesting is that despite a tiny difference in their thermodynamic stability (usually, the total energy difference is only in the order of less than 10 meV/atom[3]) their electronic properties can differ significantly. For example, the difference on band gap of SiC 4H and 3C polytypes is almost 1 eV[4] and the wurtzite form of III-V nitrides is polar whereas 3C is nonpolar, a distinction that alters profoundly the electric field profile in nitride lasers and light emitting diodes[5].

With the recent interest in the condensed matter physics community of layered two-dimensional (2D) compounds such as graphene[6], the transition metal dichalcogenides[7], and topological insulators[8], the stacking sequences of these layers is expected to take on a new importance. This is especially important in cases where the individual 2D layers exhibit structural distortions (such as inequivalent in-plane bonds), the stacking of which along the perpendicular direction creates "*natural superlattices*". Because of the great similarity in their *thermodynamic energies*, polytype physics is rather difficult to explore by conventional structural probes. Yet, various stacking sequences may maintain or break inversion symmetry that can then play an important role in keeping or lifting certain degeneracies, with implications for their electronic structure, spin polarization physics and Rashba physics. How to characterize and understand these 'electronic markers' has only been minimally addressed, either theoretically or experimentally.

As an important prototype system we focus on the layered oxides of the type $(BiS_2)/(LaO)/(BiS_2)$ where the 2D planes of $BiS_2$ are separated from each other by the LaO barrier (Fig. 1a). This material has recently received a good deal of interest because of its potential to host unconventional superconductivity up to 10.6 K[9], hidden spin polarizations[10, 11], spin field effect transistors[12] and electrically tunable Dirac cones[13], etc. In the previous studies, the compound was often assumed to have a centrosymmetric space group of P4/nmm, and to have a single specific crystallographic structure (called $T_0$ here)[14-16]. However, such high-symmetry structure having two equal in plane Bi-S bonds reported in the Inorganic Crystal Structure Database (ICSD) database has been predicted by Yildirim to be dynamically unstable[17], and further neutron diffraction experiment[18] confirmed that the two Bi-S bonds have different lengths. Some of the present authors[19] examined via density functional theory (DFT) various polytype arrangements of the individual 2D planes having unequal Bi-S bonds and predicted three stable classes of polytypes (noted as $T_1$-$T_3$), some being centrosymmetric and some breaking inversion symmetry (Fig. 1b). In this paper we show that such polytypes give rise to distinctly different symmetry-related electronic properties, even though their total energies are quasidegenerate[6]. For example, whereas in the $T_0$ structure there is a crossing of two doubly degenerate bands (so the crossing point is 4-fold degenerate, see Fig. 2a) due to its relatively higher symmetry, in other polytypes we predict two characteristic types of (partial) degeneracy removal (Fig. 2a) at X and Y points of the rectangular-shaped Brillouin zone (BZ), leading to the formation of internal "mini-gaps" within the valence band or the conduction band. In addition, the Rashba bands manifest minima at different wavevectors for different polytypes. By considering the electronic structure of different polytypes we discover certain "electronic markers" that are predicted to be sensitive to polytype stacking and the ensuing symmetry. Thus, measurements of such markers can be used in conjunction with theory, to determine structure; thereby complementing information from conventional structural probes (diffraction). Here we present detailed angle-resolved photoemission spectroscopy (ARPES) results of this system, aiming for identifying the possible polytype physics in this system. Notably, ARPES provides us with the detailed band structure to be compared with DFT calculations based on different polytypes. Our work establishes spectroscopic detection of theoretically established electronic markers as a powerful tool to delineate energetically quasi-degenerate polytypes.

**Predicted electronic markers of different LaOBiS$_2$ polytypes**

The basic layered crystal structure of LaOBiS$_2$ is shown in Fig. 1a, with a sandwiched structure containing two BiS$_2$ layers and an intermediate LaO layer. The structure (referred as "$T_0$") has high symmetry (space group #129, P4/nmm) with the x-y in-plane Bi and S atoms forming a perfect square (Fig. 1b). However, $T_0$ structure was predicted to have phonon instability via first-principle calculations[17]. Instead, an in-plane distortion causing alternation of the length of the Bi-S bonds could stabilize the structure. Considering the stacking of two BiS$_2$ layers along the z direction, one can get three polytypes by stacking layers whose Bi-S bonds are distorted along different direction (Fig. 1b) : (i) both layers distort along the x direction [(x, x), referred as polytype "$T_1$"]; (ii) one layer distorts along the x direction while the other along –x direction [(x, -x), referred as polytype "$T_2$"], and (iii) one layer distorts along the x direction while the other along y direction [(x, y), referred as polytype "$T_3$"]. All the polytypes $T_1$-$T_3$ are almost equally likely to exist in a real sample as their energy differences are quite small (~ 0.1 meV per atom[19]). The stacking direction, space group and the presence of inversion symmetry for $T_0$-$T_3$ structure are listed in Table I. Neglecting the small energy difference between polytypes, the (x, y) orientation of $T_3$ has twice the occurrence probability of $T_1$ (x, x) or $T_2$ (x, -x) but the actual mixture in a real sample could be controlled by growth effects (e.g., the interfacial energies between polytypes and growth rates).

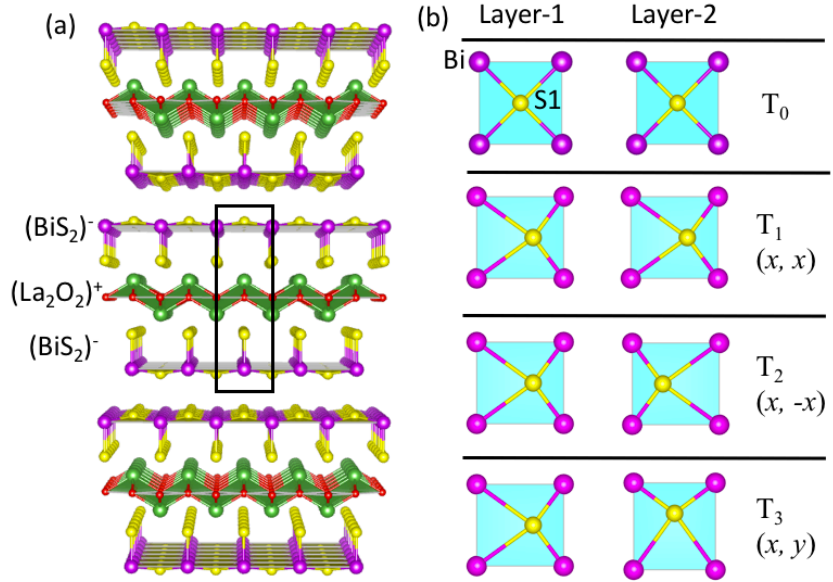

Fig: 1: (a) The layered structure of LaOBiS2 with the unit cell indicated by the black frame. The green, red, purple, and yellow balls represent La, O, Bi, and S atoms, respectively. Note that there are two BiS2 layers in each unit cell. (b) Different stacking configurations of two BiS2 layers for unstable structure T0 and its three stable polytypes $T_1$-$T_3$. Comparing with T0, the Bi-S1 2D networks of $T_1$-$T_3$ have an in-plane distortion showing the displacement of S1 atom along x or y direction, and thus two different Bi-S1 bond lengths.

We use DFT in the electronic structure calculation with the projector-augmented wave (PAW) pseudopotential[20] and the exchange and correlation of Perdew, Burke, and Ernzerfhof (PBE)[21]. The equilibrium crystal structures are obtained by DFT total energy minimization (see Methods for more details). We note the following electronic markers of polytypism:

*(i) Formation of polytype-dependent Rashba mini gap at the X or Y wavevectors in the BZ:* Among

the 4 polytypes considered here, $T_0$ has the highest symmetry with a non-symmorphic and centrosymmetric space group P4/nmm. Even though the combination of the two $BiS_2$ layers in a unit cell creates a centrosymmetric structure, the system produces a Rashba-like splitting due to the locally non-centrosymmetric nature of each $BiS_2$ sector, leading to two Rashba-like bands with opposite helical spin topology[11]. The two band crossing points at wavevectors X(Y) are superimposed by screw axis operation along the x(y) real space direction, leading to 4-fold degeneracy (including spin). The effective Hamiltonian around the wavevector X involves a 4x4 Dirac matrix, rendering a 3D Dirac cone (on a small energy scale near the degeneracy point) robust even with spin-orbit coupling (SOC)[22, 23]. On the other hand, the bands off the X(Y) wavevectors are all two-fold degenerate due to inversion symmetry and time reversal symmetry. The result is two horizontally shifted parabolas crossing at one point; we refer to this type of band structure feature as "band motif I", shown in Fig. 2a. Each polytype has two band motifs—one at X and one at Y. DFT calculation verifies that in $T_0$ polytype the BM of both X and Y valley are the same and belong to type I, as shown in Fig. 2b.

However, such Dirac points are not robust against symmetry-lowering perturbations. When the dynamically unstable $T_0$ evolves to its polytypes $T_1$-$T_3$, it looses non-symmorphic symmetries and thus removes the 4-fold degeneracy of Dirac points at certain valleys, forming internal "mini-gaps" within the valence and conduction bands . For $T_1$ (space group $Pmn2_1$) structure, the remaining symmetry operation that can protect the Dirac point is the screw axis $\{C_{2x}|(1/2, 0, 0)\}$, so it ensures Dirac cones at X, while due to the loss of $\{C_{2y}|(1/2, 0, 0)\}$ symmetry at Y point the 4-fold degeneracy splits into two Kramers pairs with a mini-gap between the splitted bands (see Fig. 2c). Thus, the X point has BM–I whereas the Y point has BM-II. On the other hand, $T_2$ (space group $P2_1/m$) has $\{C_{2y}|(1/2, 0, 0)\}$ symmetry and thus hosts Dirac points at Y and gapped states at X (see Fig. 2d). The BM having such a mini-gap at X(Y) instead of a Dirac point is "type II BM" schematically shown in Fig. 2a. Type II BM could have 2-fold degenerate bands off X(Y) due to inversion symmetry (as in $T_2$), or singly-degenerate bands off X(Y) due to the absence of inversion symmetry (as in $T_1$). Finally, $T_3$ has the lowest symmetry C2 without inversion or non-symmorphic symmetry, rendering identical gapped states at both X and Y (see Fig. 2e).

Figure 2f summarizes the BM types of X and Y valley for $T_0$-$T_3$ polytypes, showing that according to the classification by the degree of degeneracy of the high-symmetry wavevectors, each of the 4 polytypes $T_0$-$T_3$ has a unique BM(X) and BM(Y). We further note that the polytype $T_1$ that has a real-space (y, y) stacking and $T_2$ with a real space (x, -x) stacking have each one BM-I and one BM-II, so they are indistinguishable if one considers only the band motifs at the individual X and Y wavevectors. But these polytypes would be distinct if we add the marker "mini-gap" $\Delta$ for different bands and valleys, illustrating distinct markers for different polytypes. For example, for $T_1$ the mini-gap of conduction band $\Delta_{CB}$ is larger than that of the valence band $\Delta_{VB}$, while for $T_2$ the minigap $\Delta_{CB}$ is smaller than $\Delta_{VB}$. For $T_3$ the X and Y valleys are symmetric (see Table I).

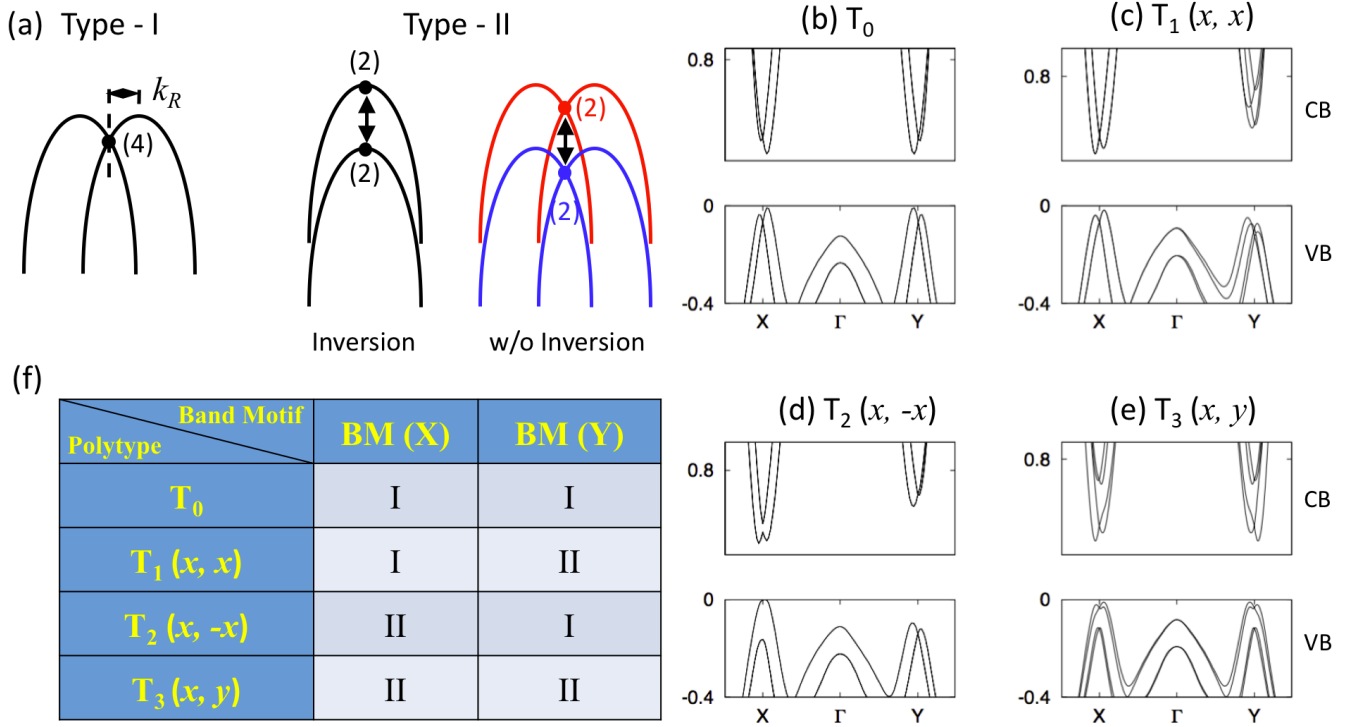

Fig. 2: (a) Illustration of two types of band motifs (BM) classified by the degree of degeneracy (indicated by the number with parenthesis) at X (1/2, 0, 0) and Y (0, 1/2, 0) points. Type-I BM manifests the 4-fold degenerate Dirac point, which breaks into a pair of 2-fold degenerate points with a mini-gap $\Delta$ (horizontal arrows) as type-II BM. Off high-symmetry points X or Y type-II BM can have 2-fold degenerate bands (black) or single-degenerate bands (red and blue) according to the presence of inversion symmetry. (b-e) Band structures of T0-T3 polytypes show different BM types at X and Y point, which act as an electronic marker. Beyond the X or Y points in either direction of these plots is the M point. The BM types are summarized in (f).

(ii) ***The minima of the Rashba bands occur in different wavevectors for different polytypes:*** The classical Rashba spin splitting manifests two parabolic band dispersions shifting towards each other in momentum space. The band edges located off the high-symmetry point are shown in Fig. 2a. In BM type

I of centrosymmetric structures ($T_0$ and $T_2$), the band splitting along G-X (G-Y) high-symmetry line that locates inside the BZ is due to the combined contribution of the hybridization between two BiS$_2$ layers and SOC, indicating a larger momentum offset. In contrast, along X-M (Y-M) located at the surface of the BZ the band splitting is purely SOC-induced, while the layer hybridization is forbidden by the non-symmophic symmetry. This effect is stronger for conduction bands in which the momentum offset along Gamma-X is 3-4 times larger than that along X-M direction, as shown in Fig. 2. On the other hand, for the non-centrosymmetric polytypes especially $T_3$, the anisotropy of momentum offset along different symmetry lines is strongly suppressed. The band edges of both conduction band and valence band of the polytypes are distinctly located at different wavevectors, as shown also in Table I. The dispersions of the conduction bands and the valence bands, especially the band edges in momentum space could be a marker for identification by ARPES spectra.

Table I: Electronic markers of various polytypes $T_0$-$T_3$ and present experimental samples of LaOBiS$_2$. For the Rashba momentum offset, the two numbers indicate $k_R$ along Γ-X(Y) and X(Y)-M.

| | | $T_0$ | $T_1$ | $T_2$ | $T_3$ | Exptl. |
|---|---|---|---|---|---|---|
| Stacking form | | --- | (x, x) | (x, -x) | (x, y) | Unknown |
| Space group | | P4/nmm | P2$_1$mn | P2$_1$/m | C2 | Unknown |
| Inversion symmetry | | Yes | No | Yes | No | No |
| Marker (i) (minigap energies) | $\Delta_{VB}(X)$ (eV) | 0 | 0 | 0.16 | 0.11 | 0.15±0.05 |
| | $\Delta_{VB}(Y)$ (eV) | 0 | 0.03 | 0 | 0.11 | 0.15±0.05 |
| | $\Delta_{CB}(X)$ (eV) | 0 | 0 | 0.03 | 0.17 | -- |
| | $\Delta_{CB}(Y)$ (eV) | 0 | 0.13 | 0 | 0.17 | -- |
| Marker (ii) | $k_{R-e}(X)$ (Å$^{-1}$) | 0.047/0.017 | 0.034/0.039 | 0.005/0 | 0.041/0.039 | 0.045±0.08 /0.045±0.08 |
| | $k_{R-h}(X)$ (Å$^{-1}$) | 0.047/0.036 | 0.043/0.029 | 0.010/0.010 | 0.056/0.028 | 0.06±0.01 /0.03±0.01 |
| | $k_{R-e}(Y)$ | 0.047/0.017 | 0.040/0.010 | 0.040/0.010 | 0.041/0.039 | 0.045±0.08 |

| Rashba momentum offsets) | ($A^{-1}$) | | | | | /0.045±0.08 |
| --- | --- | --- | --- | --- | --- | --- |
| | $k_{R-h}$(Y) ($A^{-1}$) | 0.047/0.036 | 0.068/0.038 | 0.059/0.040 | 0.056/0.028 | 0.09±0.03 /Unclear |

**Sample Growth**

The distribution of polytypes in a given sample depends on the growth protocol; the experimental determination of markers could thus depend on the sample at hand. High-quality single crystals of LaOBiS$_2$ were grown using CsCl/KCL as flux. The charge was sealed in a quartz tube, fired at 850 °C and then slowly cooled down to room temperature. The technique is similar to that described in literature[24]. The size of the single crystals studied is of order 2.0 x 2.0 x 0.3 mm$^3$. Chemical compositions of the single-crystals were determined using energy dispersive X-ray analysis (EDX) (Hitachi/Oxford 3000) and single-crystal X-ray diffraction. The crystal structure of the single-crystals was determined using a Rigaku X-ray diffractometer XtaLAB PRO equipped with PILATUS 200K hybrid pixel array detector at the Oak Ridge National Laboratory.

**Band structure, dispersion and Fermi surface - electronic markers from ARPES**

Figure 3a shows the ARPES measured Fermi map in the first Brillouin zone, with four small electron pockets found at the X/Y points, which agree with the theoretical predictions as well as previous measurements[25, 26]. Figure 3b shows an example of the zoomed-in spectra on one of the electron pockets, with decreasing intensity in the second Brillouin zone. The Fermi surface around the X, Y points form two contour loops with a nearly square shape, indicating the Rashba band splitting. The band dispersions for the conduction bands and valence bands along the high symmetry cut Γ-Y and X-M direction are shown in Figure 3c. From these we see that the band gap between the CB and VB is about 0.9 eV and that the sample is lightly doped n-type. This doping allows a clear view of both the conduction and valence bands with ARPES.

Examples of the zoomed-in spectra of the conduction bands and valence bands along the high symmetry direction M-X-M and Y-Γ-Y are presented in Fig. 4, to be compared with DFT predictions of T$_0$, T$_1$, T$_2$ and T$_3$ respectively. A few observations are made in the following: (1) the typically assumed structure (single T$_0$ polytpe) is not able to explain the data; (2) the data supports a superposition of polytpes as no single polytpe can explain all the data; and (3) the dominant polytpe appears to be T$_3$ type as this captures the majority of the major features.

For the CB spectrum along M-X-M the experimental data (top left) shows a clear internal or central state that is absent in the calculated spectrum for $T_0$, as marked by the arrow in the 2nd panel down, left column. This is a clear marker that we need to go beyond the simplest structure ($T_0$ polytype). The other polytypes $T_1$, $T_2$, and $T_3$ as calculated for this cut all show the central state, though the differences between the calculated structures for this cut are similar enough that we should look to other cuts to distinguish between these possibilities.

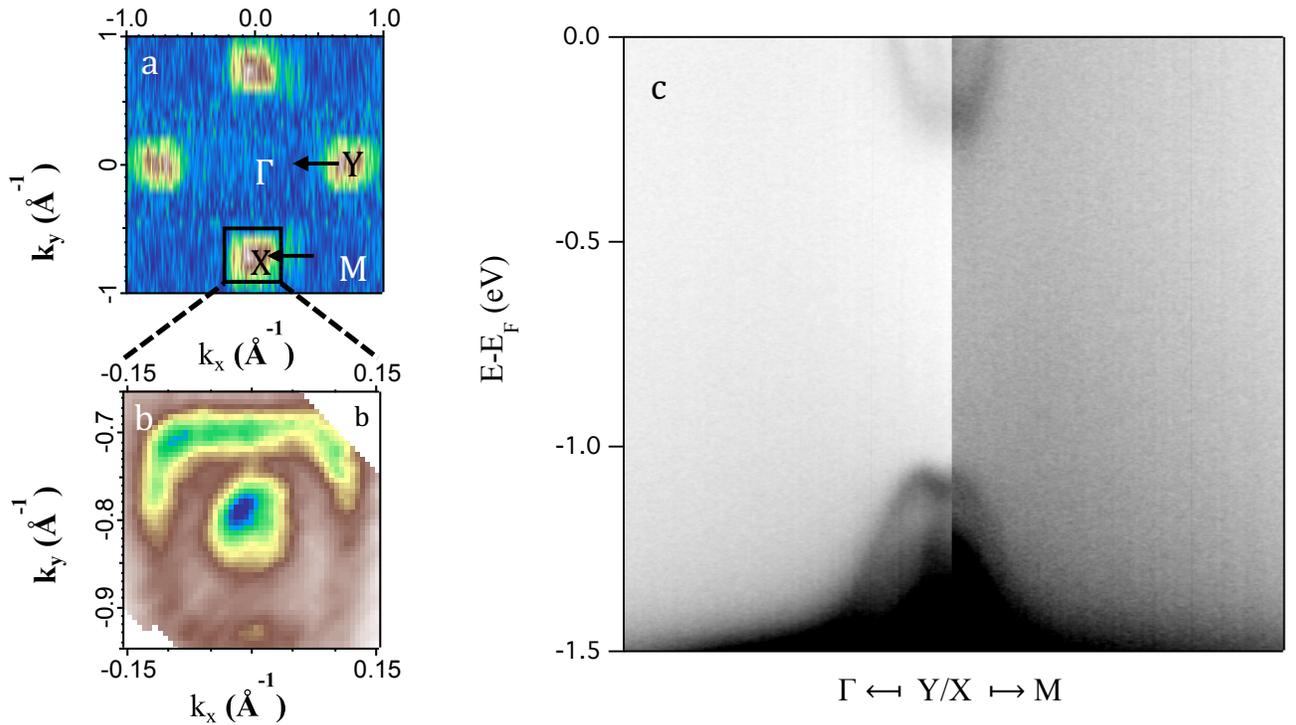

Figure 3 a) Fermi map of lightly n-doped $LaBiOS_2$ in the first Brillouin zone. Electron pockets are found near the X/Y points, the zoomed-in version of which is shown in b). ARPES spectra along high symmetry cut Γ-Y/X-M (black arrows) are shown in c) for conduction bands and valence band. The data is raw and unsymmetrized.

On the other hand, the experimental conduction band minima shown in Figs. 4a and 4b are at very much the same energy, which is at odds with the theoretical prediction of a significant anisotropy of the CB minimum for both the $T_1$ and $T_2$ polytypes, as highlighted in columns a and b. We note that the energy difference of the CB minima for polytpes $T_1$ and $T_2$ is a few hundred meV different than for T0 and T3, even though the overall system energy Therefore, $T_1$ and $T_2$ can be ruled out as the lone or dominant polytypes, though a superposition of the two of these (minidomains) or the addition of these with other polytpes could explain the lack of X/Y anisotropy of the CB minima.

Next, we consider more subtle but still clear effects in the data, especially the Rashba 'minigaps', wich are a deviation from the 'classical' type of Rashba splitting that is well-known for many materials. This Rashba minigap is most clear in the experimental panel c, with it being more filled in or fully absent for the other experimental cuts. For the theoretical cuts corresponding to panel c, we see that only polytypes $T_2$ and $T_3$ show the minigap, implying that at least one of these should have reasonably strong spectral weight. The experimental mini gap magnitude is estimated to be 0.15±0.05 eV between the valence bands, consistent with the predictions of DFT on the $T_2$ and $T_3$ structures.

ARPES spectra along the Γ-Y cut, as shown in Fig. 4d, shows a suppression of spectral weight in the regime of the minigap, but there are also clearly states there as well. This would seem to favor a superposition of $T_3$ with any of $T_0$, $T_1$, $T_2$, all of which have the presence of "central states". The experimental momentum offsets (Marker II) are also tabulated in Table I, and are also most consistent wit the $T_3$ structure as the dominant polytype. The data as a whole indicates that the $T_3$ structure contributes most significantly to the ARPES signal, with an additional $T_1/T_2$ mixing as a secondary effect. A similar conclusion has been recently reached from the latest X-ray and neutron scattering results using very different but complementary metrics[27].

**Conclusion**

In conclusion, in this work we demonstrate the capability of ARPES to distinguish subtle electronic markers of different polytypes in the layered compound LaOBiS$_2$. We found the ARPES spectra are mostly consistent with the system being largely the $T_3$ polytype, the knowledge of which might be crucial to our understanding of the novel properties in this system. While this work focused on a specific material, it demonstrated that different polytypes or a mixture of those can be generally identified by their relevant "electronic markers" with ARPES techniques. In the future, optical or transport experiments such as *interband* absorption and emission may also be utilized to observe or take advantage of such features, especially if the Fermi energy could be tuned to the middle of some of these Rashba minigaps.

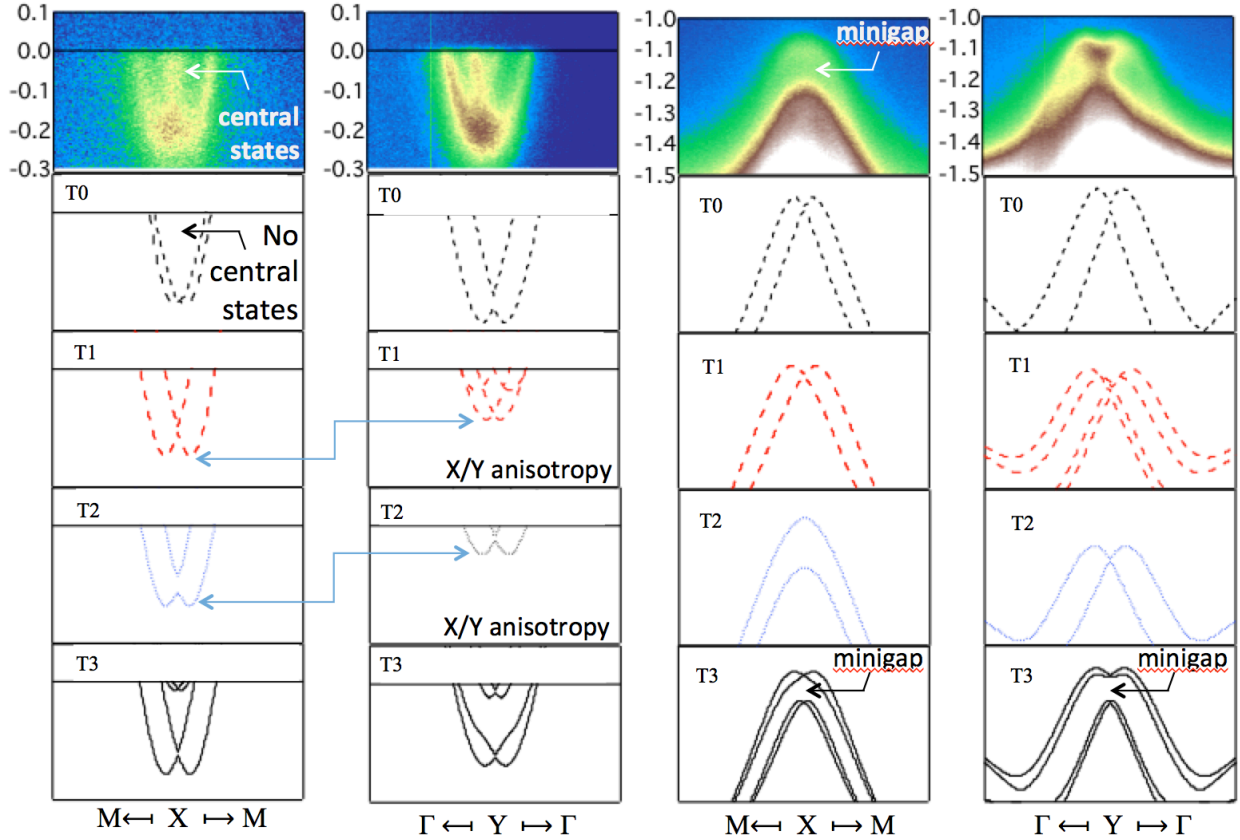

Figure 4 Experimental spectra of a) conduction bands along M-X-M direction; b) conduction bands along Y-Γ-Y direction c) valence bands along M-X-M direction and d) valence bands along Y-Γ-Y direction, to be compared with DFT calculations of T0, T1, T2 and T3 respectively.

**Methods**

The equilibrium crystal structure was obtained by DFT total energy minimization performed with an energy tolerance of $10^{-4}$ eV, and all atomic positions were relaxed with a force tolerance of $10^{-3}$ eV/Å. The electronic structures were calculated by using the projector-augmented wave (PAW) pseudopotential[20] and the exchange and correlation of Perdew, Burke, and Ernzerfhof (PBE)[21] as implemented in the Vienna *ab initio* package (VASP)[28]. The plane wave energy cutoff (reflecting basis set size) was set to 550 eV. Spin-orbit coupling was included as a perturbation to the pseudopotential throughout the calculation. We note that he PBE functional didn't take the long-range van der Waals (VDW) interaction into account, and thus usually overestimate the interlayer space. However, the VDW correction underestimates the in-plane lattice constant and thus causes the $T_0$ structure to be the ground state, which contradicts the theoretical prediction and experiments. Since the accuracy of the in-plane lattice constant is more important to investigate the polytype physics, we used PBE functional for the base of calculation and analysis.

**Acknowledgements**

This work was funded by NSF DMREF project DMR-1334170 to the University of Colorado and the University of Kentucky. We thank Feng Ye for help with some of the early structural characterizations of these materials. The Advanced Light Source is supported by the Director, Office of Science, Office of Basic Energy Sciences, of the U.S. Department of Energy under Contract No. DE-AC02-05CH11231.